\documentclass{article}%
\usepackage{amsmath}
\usepackage{amsfonts}
\usepackage{amssymb}
\usepackage{graphicx}%
\providecommand{\U}[1]{\protect\rule{.1in}{.1in}}

\begin{document}

\title{On the use of DMT approximations in adhesive contacts, with remarks on random
rough contacts}
\author{M.Ciavarella\\Politecnico di BARI. \\V.le Gentile 182, 70125 Bari-Italy. \\Email mciava@poliba.it}
\maketitle

\begin{abstract}
The contact between rough surfaces with adhesion is an extremely difficult
problem, and the approximation of the DMT theory (to neglect deformations due
to attractive forces), originally developed for spherical contact of very
small radius, are receiving some new interest. The DMT approximation leads to
extremely large overestimate of the adhesive forces in the case of spherical
contact, except at pull-off. For cylindrical contact, the opposite trend is
found for larger contact areas. These findings suggest some caution in solving
rough contacts with DMT models, unless the Tabor parameter is really low.
Further approximate models like that of Pastewka \& Robbins' may be explained
to work only for a coincidence of error cancellation in their range of parameters.

\end{abstract}

Keywords: Adhesion, Maugis' theory, rough surfaces, DMT theory, JKR theory

\bigskip

\section{Introduction}

The Derjaguin-Muller-Toporov (DMT) theory (Derjaguin et al., 1975, Muller et
al., 1980, 1983), for the contact of elastic spheres with adhesion, has a long
history. After Bradley (1932) and Derjaguin (1934) obtained the adhesive force
between two \textit{rigid} spheres, equal to $2\pi Rw$, where $w$ is the work
of adhesion, and $R$ is the radius of the sphere, JKR (Johnson Kendall and
Roberts 1971) developed a theory for elastic spheres, assuming adhesive forces
occur entirely within the contact area, obtaining $3/4$ of the Bradley
pull-off value, and hence the independence on the elastic modulus raised a
long debate about the a comparison of the pull-off prefactor. 

As the main attention in the diatribe between JKR and DMT \bigskip was limited
to the pull-off value, it is often believed that DMT is the limit for Tabor
parameter (Tabor, 1977)\bigskip%
\begin{equation}
\mu=\left(  \frac{Rw^{2}}{E^{\ast2}\Delta r^{3}}\right)  ^{1/3}=\frac{\left(
Rl_{a}^{2}\right)  ^{1/3}}{\Delta r}=\frac{\sigma_{th}}{E^{\ast}}\left(
\frac{R}{l_{a}}\right)  ^{1/3}\rightarrow0\label{Tabor}%
\end{equation}
where $\Delta r$ is the range of attraction of adhesive forces, close to
atomic distance, and $E^{\ast}$ the plane strain elastic modulus. Also, we
have introduced the length $l_{a}=w/E^{\ast}$ as an alternative measure of
adhesion, and $\sigma_{th}$ is the theoretical strength of the material. Now,
while it is true that DMT predicts the Bradley result for the force at
pull-off also for elastic spheres, the DMT theories have been much less
compared with exact results, when considering the entire load-displacement
curves. In both DMT methods,

\begin{itemize}
\item the attraction forces act exclusively \textit{outside} the contact, and

\item the repulsive forces \textit{only} are responsible for deformation.
\end{itemize}

Then, in the DMT "force method":-

\begin{itemize}
\item the force of adhesion can be simply obtained by integrating, according
to Derjaguin's approximation, the forces of facing elements outside of the
contact, separated by a gap which is given by Hertz theory.
\end{itemize}

We shall concentrate on the latter (force) method, which is what is commonly
used when DMT approximation is considered in the generalized context of rough
contact (see Persson \& Scaraggi, 2014). In one looks at the force of adhesion
\textit{not at pull-off, }it decreases from $2\pi Rw$ to $\pi Rw$, in the
"thermodynamic method"\footnote{In the "thermodynamic method", the force is
computed by the rate of change of surface energy as the sphere is pressed with
approach $\alpha$, i.e. $dW_{s}/d\alpha$. It turns out that the
\textquotedblleft thermodynamic\textquotedblright\ method tends to give
opposite error with respect to the "force method", and it is also more
complicated to use, so it has not received much attention. } while it
increases in the "force method", as shown by (Muller et al., 1983), and
Pashley (1984). Pashley (1984) in particular notices that in the force method,
the adhesive force should be always larger than $2\pi Rw$, the value obtained
for a truncated rigid sphere independently on the contact radius, as the
Hertzian profile is closer to the flat surface than the rigid spherical profile.

Maugis (2000), in his Maugis-Dugdale analysis (which \textit{does not }make
the DMT approximations) called the low $\mu$ end the "DMT theory", which in
fact is now the version most commonly associated with DMT, and sometimes
called DMT-M. In this version, the attractive forces are constant, and equal
to the pull-off value, $2\pi Rw$. This is indeed what comes out from DMT
theory, but only in the limit of $\mu=0$: therefore, DMT is exact only in this
limit case, and for any finite value of $\mu$, DMT theories give an error
which we shall estimate in fact in details in the present paper as a function
of the Tabor parameter, since the previous estimates of (Muller et al., 1983),
and Pashley (1984) do not clarify clearly the role of Tabor parameter.
\bigskip Greenwood (2007) also has discussed more details of the DMT theory in
the limit $\mu\rightarrow0$.

But we shall not limit ourselves to the spherical contact case, since this
case has been given already much attention, and is only one special case. The
DMT approximation is gaining relevance more recently again, in the context of
rough contact, where there is a lot of interest in simplifying the problem
since the JKR assumption leads to very complicated and hysteretic behaviour,
which so far, has not been included in a framework of any theory, despite some
attempts (Persson, 2002, Ciavarella, 2015). Moreover, as roughness at the
small scales seems to point to low values of Tabor parameter, the "almost
rigid" behaviour has some fundamental interest. Persson \& Scaraggi (2014)
have indeed attempted using the DMT approximations using the Persson's theory
for adhesionless contact, and seemed to find some reasonable accuracy at least
for the range of parameters they observed. Also, Pastewka \& Robbins (2014, PR
in the following) make some scaling predictions which seem to fit well some
limited range of their extensive full numerical simulations involving
atomistics rough solids. We made a first attempt to discuss PR findings in
Ciavarella (2016) where we noticed that, if PR were concerned with spherical
contact, using the DMT approximation with the additional simplification of
using only the asymptotic first term in the expansion of the gap outside the
Hertzian contacts, they would find easily large errors. But one limit to this
estimate is that we assumed circular contact, whereas PR calculation shows
more like 2D fractal contact area, perhaps closer to very elongated contacts
like in 2D cylindrical contact --- indeed, as we will discuss below, they find
a characteristic diameter of the contact independent on load, and load only
affects the elongation of the contact area. Therefore, in the present note we
develop a simple 2D line contact DMT model, we give more details about the DMT
limit for the sphere, and make further comparisons with the DMT rough contact results.

\section{\bigskip A 2D DMT-Maugis line contact model}

For 2D contact with "repulsive" diameter $d_{rep}=2a$, the full form of the
gap outside the contact is
\begin{equation}
\frac{h\left(  c\right)  }{a}=\frac{a}{R}f\left(  \frac{c}{a}\right)
\label{full-gap}%
\end{equation}
where $c>a$ and (Johnson and Greenwood, 2008)
\begin{equation}
f\left(  \frac{c}{a}\right)  =\frac{1}{2}\left[  \frac{c}{a}\sqrt{\left(
\frac{c}{a}\right)  ^{2}-1}-\log\left[  \frac{c}{a}-\sqrt{\left(  \frac{c}%
{a}\right)  ^{2}-1}\right]  \right]
\end{equation}
whose first term in the series expansion near $c=a$ is $f_{as}\left(  \frac
{c}{a}\right)  =\frac{\sqrt{8}}{3}\frac{a}{R}\left(  \frac{c}{a}-1\right)
^{3/2}$, is used in the PR version of DMT method, as commented in (Ciavarella,
2016), and in the later paragraph. We shall assume for the potential, a Maugis
simple law. This will permit a direct comparison with the "exact" Maugis
solution including deformations induced by the adhesive stresses, given by
Baney and Hui (1997) Morrow and Lovell (2005) and Johnson and Greenwood
(2008), whereas Jin et al. (2014) give a double Hertz solution which show that
results will not differ much with those with other choices of potential.

The pull-off force is not a simple multiple of $Rw$ as with circular contacts,
but varies from $P_{rigid}=\sqrt{8R\sigma_{th}w}$ to $P_{JKR}=\frac{3}%
{4}\left(  4\pi E^{\ast}Rw^{2}\right)  ^{1/3}=\frac{3}{4^{2/3}}\left(  \pi
E^{\ast}Rw^{2}\right)  ^{1/3}$, so it depends on elastic modulus.

\bigskip We define the following non-dimensional contact radius and load
\begin{align}
a^{\ast}  &  =\frac{a}{2\pi^{1/3}R^{2/3}l_{a}^{1/3}}\label{adim}\\
P^{\ast}  &  =\frac{P}{\left(  \pi E^{\ast}Rw^{2}\right)  ^{1/3}}=\frac
{P}{4^{2/3}P_{JKR}/3} \label{Pdim}%
\end{align}
and accordingly the Hertz and JKR limits are found as (Johnson and Greenwood,
2008)
\begin{align}
P_{Hertz}^{\ast}  &  =a^{\ast2}\\
P_{JKR}^{\ast}  &  =a^{\ast2}-2\sqrt{a^{\ast}}=P_{Hertz}^{\ast}-2\sqrt
{a^{\ast}}%
\end{align}
whereas the Maugis-Dugdale model shows a smooth transition between the Hertz
and JKR limit --- unlike the 3D case, where there is a transition from Bradley
rigid to JKR model. Notice that the rigid limit is subtle: while there is a
tendency to the Hertz regime, the actual pull-off in rigid limit is not zero.

Moving to a DMT force method estimate, setting the gap to $\Delta r$ gives
\begin{equation}
f\left(  \frac{c}{a}\right)  =\frac{R\Delta r}{a^{2}} \label{full-equation}%
\end{equation}
and it is clear that the approximation is good until $\frac{R\Delta r}{a^{2}%
}<1$. In dimensionless notation, $R\Delta r/a^{2}=\left(  4\pi^{2/3}a^{\ast
2}\mu\right)  ^{-1}.$

Using the asymptotic term, the lateral distance defining the size of
attractive region (which is composed of two strips of size $d_{att}$) is%
\begin{equation}
d_{att,asym}=\left[  \frac{\left(  \frac{3}{2}R\Delta r\right)  ^{2}}{d_{rep}%
}\right]  ^{1/3} \label{datt}%
\end{equation}
and when contact radius is small, we require a correction from the solution of
(\ref{full-equation}), $\frac{d_{att}}{a}=\frac{c}{a}-1=\beta\frac
{d_{att,asym}}{a}$. As we are using the Maugis potential, the attractive load
is therefore simply the product of the theoretical strength and the area of
the adhesive strips, $P_{DMT,att}=2d_{att}\frac{w}{\Delta r}$. Using
(\ref{datt}), the attractive load is obtained as%
\begin{equation}
P_{DMT,att}^{3}=-2^{3}\beta^{3}d_{att}^{3}\left(  \frac{w}{\Delta r}\right)
^{3}=-2^{3}\beta^{3}\left(  \frac{3}{2}\right)  ^{2}R^{2}\Delta r^{2}\frac
{1}{2a}\left(  \frac{w}{\Delta r}\right)  ^{3}%
\end{equation}
Using (\ref{adim}, \ref{Pdim}) and Tabor parameter (\ref{Tabor}),%
\begin{equation}
P_{DMT,att}^{\ast}\simeq-\beta\left(  \frac{\mu}{a^{\ast}}\right)  ^{1/3}%
\end{equation}

The results of the DMT theory are presented in Fig.1a (dashed lines) for
$\mu=0.05,0.25,1,5$, together with the Maugis solution of Johnson and
Greenwood (2008) which we take as reference as "exact". In Fig.1b, we compare
the JG Maugis solution with the further approximation of taking only the first
term in the gap profile, which clearly leads to serious errors even at low
Tabor parameter. We shall explain why in the PR use of this further
approximation, the error was probably balanced by another approximation.

\begin{center}
$%
\begin{array}
[c]{cc}%
{\includegraphics[
height=2.9231in,
width=4.4105in
]%
{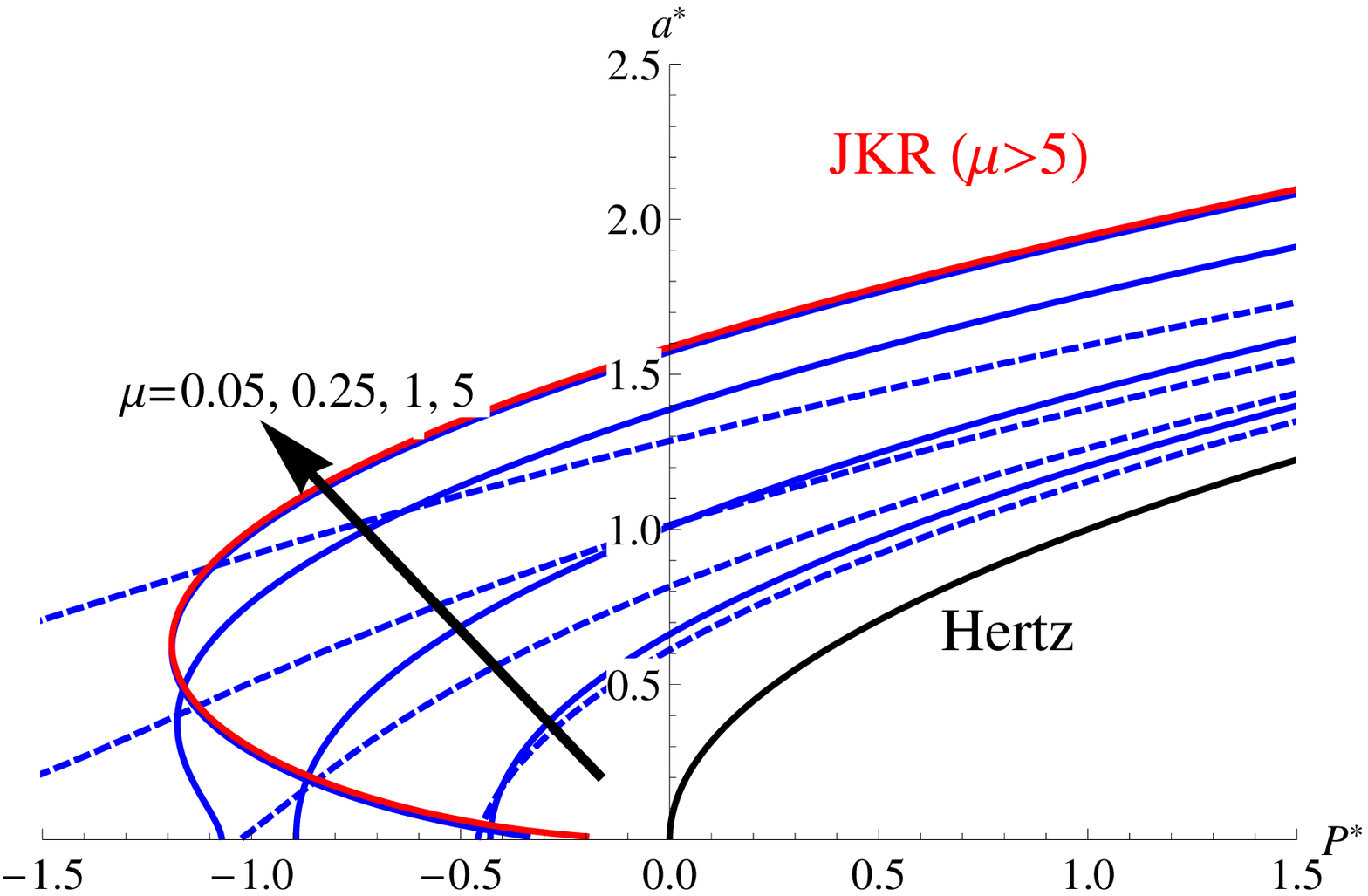}%
}
& (a)\\%
{\includegraphics[
height=3.0078in,
width=4.5282in
]%
{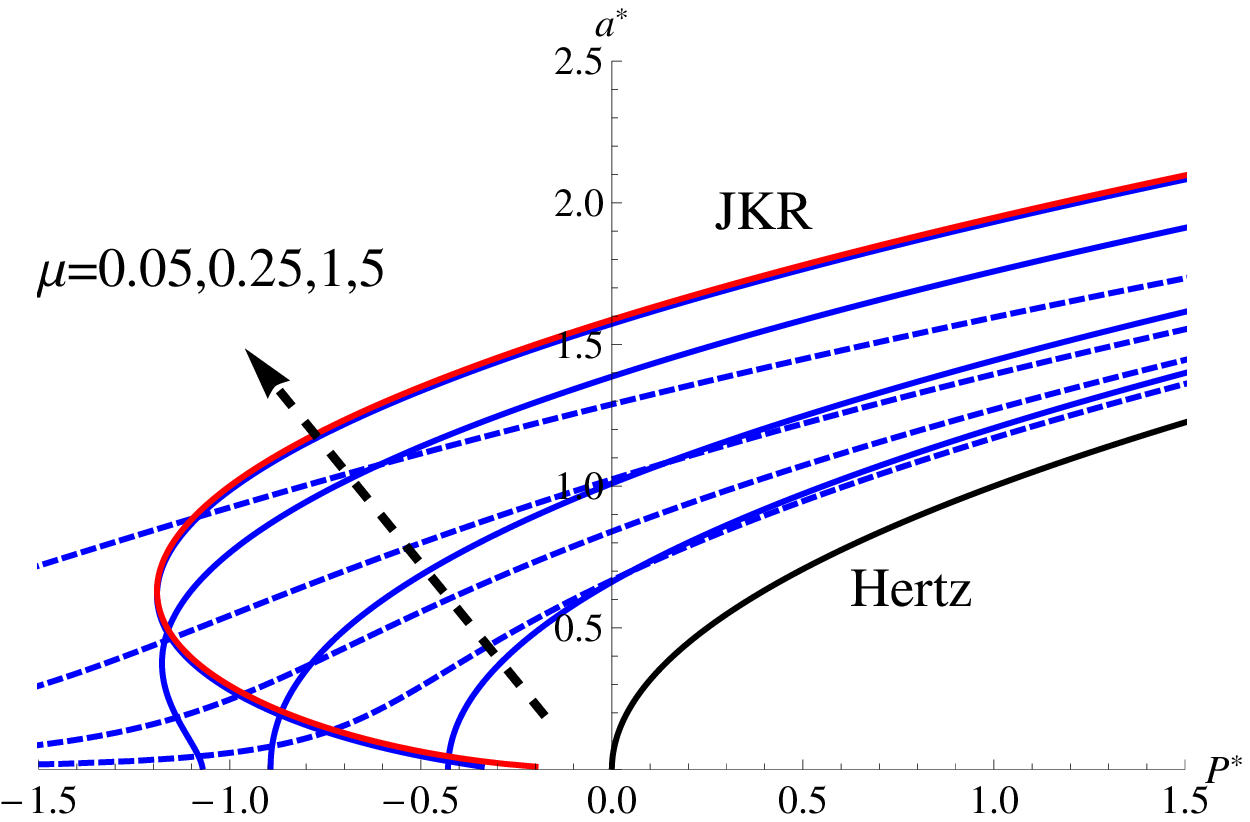}%
}
&
\end{array}
$

Fig.1 Line contact. Comparison of Maugis (a) "exact" solution (solid blue
line), with the DMT model with (a) full shape of the gap (dashed line), or
with the first term of the gap (b). The black (red) solid line are Hertz (JKR)
limits. The arrow shows increasing values of Tabor parameter $\mu
=0.05,0.25,1,5$.
\end{center}

It is clear that the DMT theory gives a reasonable result only for $\mu<0.05$
as at $\mu=0.25$, the error is already significant, of the order of 20\% at
pull-off. Errors become large at $\mu>1$, particularly at pull-off, as larger
than 100\%. This is clearer from fig.2, where the pull-off values are plotted.

\begin{center}
$%
\begin{array}
[c]{cc}%
\raisebox{-0pt}{\parbox[b]{4.0274in}{\begin{center}
\includegraphics[
height=2.6775in,
width=4.0274in
]%
{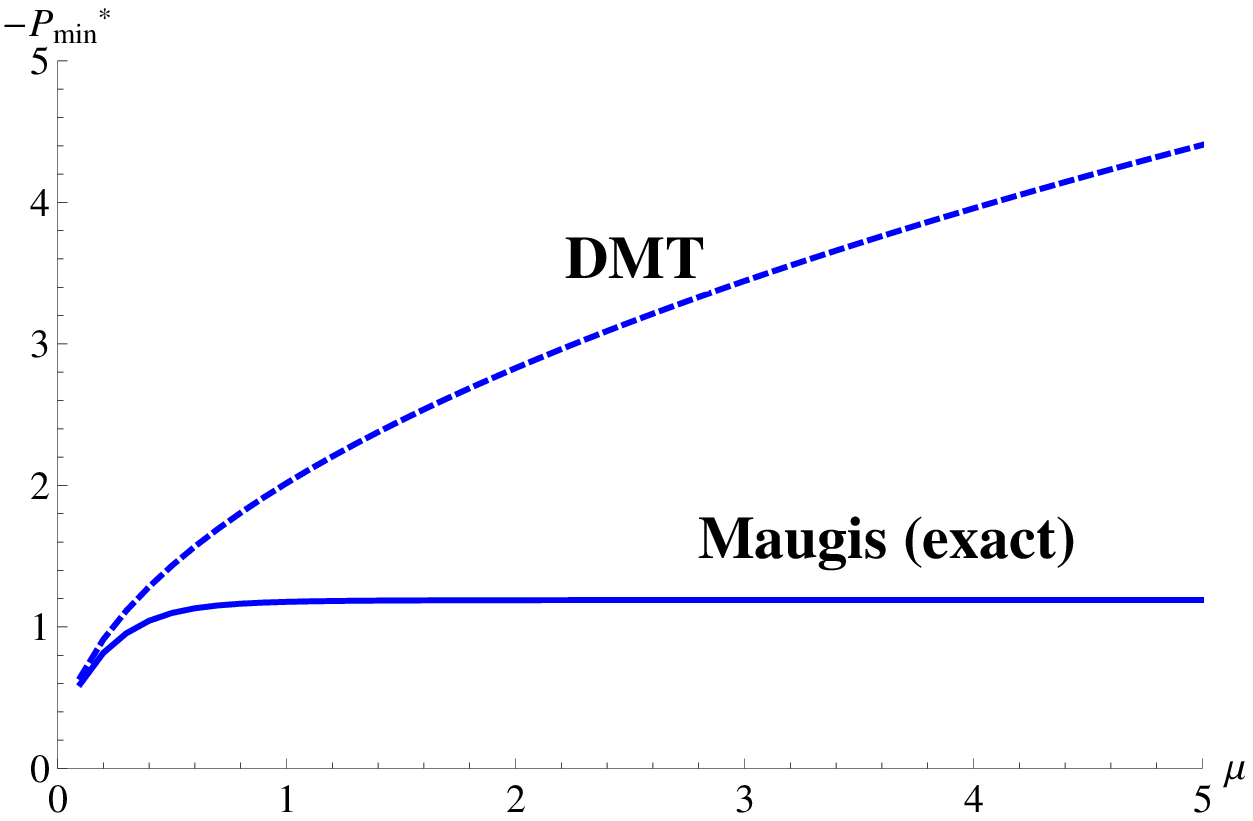}%
\\
{}%
\end{center}}}
&
\end{array}
$

Fig.2 Line contact. Comparison of pull-off values for Maugis "exact" solution
(solid blue line), with the DMT model with full shape of the gap (dashed
line). 
\end{center}

\section{\bigskip The spherical DMT model}

As this case is classical, DMT has been compared with JKR and other models in
a large number of papers. However, the comparison is mostly done for the value
at pull-off, where of course the DMT model gives the correct Bradley result
for $\mu=0$: only perhaps (Muller et al., 1983), Pashley (1984), and Greenwood
(2007) discuss more details of DMT at compressions larger than zero, and even
they, do not fully clarify the error as a function of Tabor parameter.

We shall assume for the potential, a Maugis simple law, consistently to the
line contact of the previous paragraph. We have already discussed a further
simplified form of this model in (Ciavarella, 2016) inspired by PR paper
(Pastewka \&\ Robbins, 2014), namely using the first term asymptotic form of
the gap function outside of the contact, and of computing the area of
attraction by multiplying the perimeter of the contact by a $d_{att}$ size of
attractive region. As we already discussed the very strong effect of both
these approximations on the DMT solution, we shall try here to give a fair
assessment of a full\ DMT model, and we do not make such further assumptions.

Therefore, we introduce the following dimensionless notation for the approach,
and the load in spherical contact%
\begin{equation}
\widehat{\delta}=\delta/\left(  \mu\Delta r\right)  \quad;\quad\widehat
{P}=P/\left(  \pi Rw\right)  \label{normalization}%
\end{equation}

We know the expression of the gap outside the contact area (Muller et al.,
1983), Pashley (1984), and Greenwood (2007)
\begin{equation}
\frac{h\left(  \frac{r}{a}\right)  }{a}=\frac{a}{\pi R}\left[  \sqrt{\left(
\frac{r}{a}\right)  ^{2}-1}-\left(  2-\left(  \frac{r}{a}\right)  ^{2}\right)
\arctan\sqrt{\frac{r^{2}}{a^{2}}-1}\right]
\end{equation}
Imposing the gap is equal to the range of attractive forces, gives the size of
attractive region. Writing the size of the region using the first order
expansion of the gap is possible in closed form, giving $d_{att,asym}%
=\frac{d_{rep}}{2}\left(  \frac{3\pi R\Delta r}{2\sqrt{2}d_{rep}^{2}}\right)
^{2/3},$ and the correct solution with the full expression of the gap requires
a numerical correction for small contact areas, which is $d_{att}=\beta
_{s}d_{att,asym}$. We write $\beta_{s}$ to avoid confusion with the equivalent
coefficient in line contact. For the circular geometry Hertz theory gives
$\delta=d_{rep}^{2}/\left(  4R\right)  $ and hence, using the notation of PR,
$d_{rep}$ for the repulsive contact diameter, and $d_{att}$ for the size of
attractive region (not a diameter), we have
\[
P_{att}=-\frac{\pi}{4}\left[  \left(  d_{rep}+2\beta_{s}d_{att}\right)
^{2}-d_{rep}^{2}\right]  \frac{w}{\Delta r}%
\]

Normalizing with (\ref{normalization}), we have%
\begin{equation}
\widehat{P}_{att}=-\mu\widehat{\delta}\left[  \left(  1+\beta_{s}\left(
\frac{3\pi}{8\sqrt{2}\mu\widehat{\delta}}\right)  ^{2/3}\right)
^{2}-1\right]
\end{equation}

The result can be compared with the Maugis solution for $\mu\rightarrow0$
(which indeed is the "DMT" limit for $\mu\rightarrow0$), and the JKR theory
(Johnson, et al., 1971)\footnote{JKR is presented in a curve fitted form in
order to be easily used.}
\begin{align}
\widehat{P}_{\mu=0}  &  =\frac{4}{3\pi}\widehat{\delta}^{3/2}-2\\
\widehat{P}_{DMT}  &  =\frac{4}{3\pi}\widehat{\delta}^{3/2}-\left(  \frac
{1}{4}\left(  \frac{3\pi}{8\sqrt{2}\mu\widehat{\delta}}\right)  ^{4/3}%
-1\right)  \beta_{s}\widehat{\delta}\mu\label{PR-dimensionless}\\
\widehat{P}_{JKR}  &  =\widehat{P}_{0}-1.1\left(  \widehat{\delta}%
-\widehat{\delta}_{0}\right)  ^{1/2}+0.43\left(  \widehat{\delta}%
-\widehat{\delta}_{0}\right)  ^{3/2}%
\end{align}
where $\widehat{\delta}_{0}=-\frac{3}{4}\pi^{2/3},\widehat{P}_{0}=-5/6$ are
the JKR values at pull-off in displacement control.\ 

Fig.3 reports the DMT solution, with the correct Maugis solution for various
values of $\mu$, and immediately it appears that \textit{all of the DMT
solutions are incorrectly below the }limit for $\mu\rightarrow0$, while the
correct Maugis solutions are all above this limit --- indeed, already at
$\mu=0.05$ the Maugis solution is quite distinct from the limit solution
$\mu\rightarrow0$ (which, adding to the confusion, Maugis calls the
DMT\ limit!). Not shown is also that the correction for the full form of the
gap is necessary at small indentations, as otherwise there is an additional
spurious increase of adhesive forces estimate.

Fig.4 reports the values of the pull-off estimate with the DMT method, as
compared with the Maugis solution.

\begin{center}
$%
\begin{array}
[c]{cc}%
\raisebox{-0pt}{\parbox[b]{5.0548in}{\begin{center}
\includegraphics[
height=3.2456in,
width=5.0548in
]%
{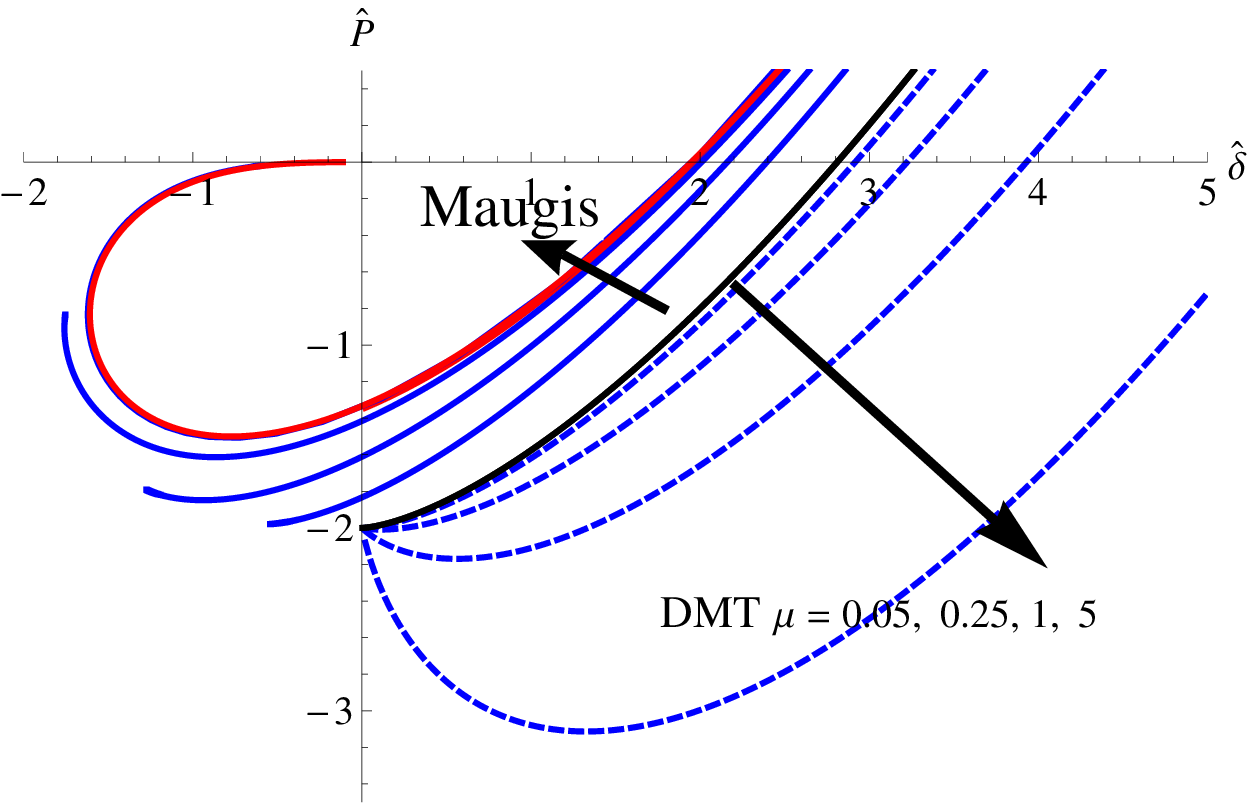}%
\\
{}%
\end{center}}}
&
\end{array}
$

Fig.3 Spherical contact. Comparison of Maugis "exact" solution (solid blue
line), with the DMT model with full shape of the gap (dashed line). The black
\ and red solid line are $\mu=0$ (what Maugis calls the DMT limit), and JKR
limits. The arrow shows increasing values of Tabor parameter $\mu
=0.05,0.25,1,5$.

$%
\begin{array}
[c]{cc}%
\raisebox{-0pt}{\parbox[b]{4.0274in}{\begin{center}
\includegraphics[
height=2.6887in,
width=4.0274in
]%
{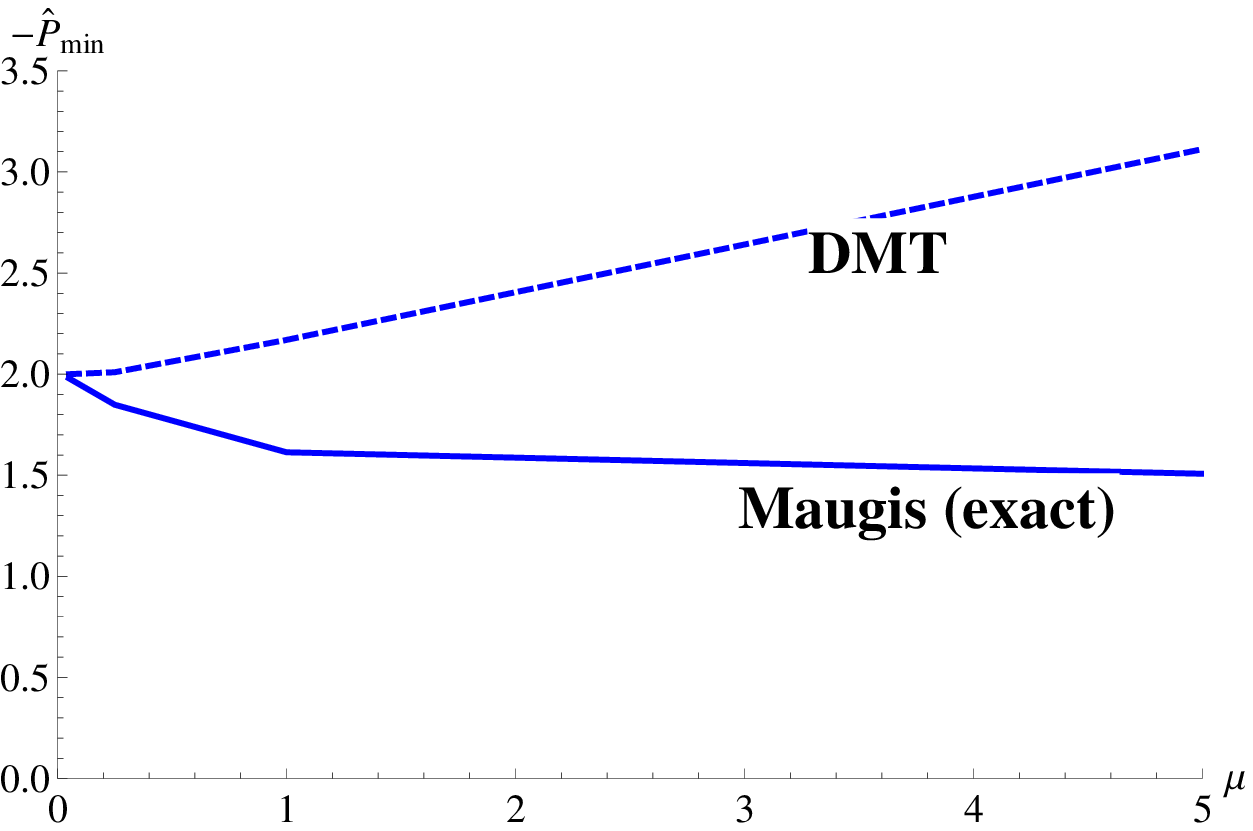}%
\\
{}%
\end{center}}}
&
\end{array}
$

Fig.4. Spherical contact. Comparison of pull-off values for Maugis "exact"
solution (solid blue line), with the DMT model with full shape of the gap
(dashed line). 
\end{center}

\section{DMT for rough contacts}

\bigskip These simple estimates suggest that the order of error in using the
DMT approximation in actually solving rough adhesive contact problems, like
done by Persson \& Scaraggi (2014) for example, could be very large, unless
Tabor parameter is quite low, and the body very close to "rigid".
Incidentally, this means a condition stricter than just Tabor parameter lower
than, say, 0.2 at the smallest scale in the model, since at the large scale,
the contact may result hysteretic, and the DMT model would lead to very large
errors. This may be in practice a very strong condition and would therefore
limit significantly the use of such methodology. In the particular case, the
Persson-Scaraggi method, which relies on use of additional approximations in
solving the adhesiveless contact and integrating for the adhesive force,
doesn't lead to simple analytical results so it is also important to check, in
the end, if there is a significant advantage with respect to a full numerical
solution of the problem, like done by Pastewka and Robbins (2014).

Speaking of PR, they interpreted their "exact" results with some very simple
estimates based on DMT approximation, and further simplifications. One may
wonder why PR obtained a reasonable fit of their data, with assuming first
order expansion of the gap function, which for example would lead to
completely wrong results for very small contact area (see Fig.1b). In
particular, this is surprising since their Tabor parameter are said to be of
the order of 1, and  therefore not very small.

PR, in trying to find some scaling equations, suggested to use a DMT-like
force method calculation of the attractive force during the loading stage of
what they call "non-sticky" cases, based on an asymptotic first term for of
the gap,
\begin{equation}
2\frac{h\left(  x\right)  }{d_{rep}}=\frac{\sqrt{8}}{3}h_{rms}^{\prime}\left(
\frac{2x}{d_{rep}}\right)  ^{3/2}\label{PR-lateral}%
\end{equation}
where $h_{rms}^{\prime}$ is \textit{not }the slope at the contact edge
obviously equal to $a/R$, as in a standard Hertzian contact, \textit{but a
slope value} they say estimated from the random roughness, and so anyway it is
a fixed geometric quantity. This apparently innocuous assumption hides a very
important effect: while taking the first order in the gap leads to very large
error (see Fig.1b) at small areas, here PR take a representative value which
can only permit to fit results of their cases --- as they have $h_{rms}%
^{\prime}=0.1$ or $0.3$ this corresponds to a DMT model at $a/R=0.1,0.3$
respectively. If they had much smaller $h_{rms}^{\prime}$ or much higher (as
indeed it is very possible), then we would not know the order of the error.
Therefore, we suspect that it is a pure coincidence that they were able to
make reasonable fits.

However, since the rough contact is such that they find (numerically) an
average representative diameter of the elongated areas of contact of
\begin{equation}
d_{rep}=4h_{rms}^{\prime}/h_{rms}^{\prime\prime}=2h_{rms}^{\prime
}R\label{drep}%
\end{equation}
where $h_{rms}^{\prime},h_{rms}^{\prime\prime}$ are rms slopes and curvatures,
and we used $R=2/h_{rms}^{\prime\prime}$.\ Hence, $a/R=h_{rms}^{\prime}$, and
it is not just an estimate that PR do when exchanging $h_{rms}^{\prime}$ with
$a/R$ in (\ref{PR-lateral}), but a quite correct substitution. Their contacts
have elongated shapes whose representative size is of the order of the
smallest wavelength. Indeed, we can estimate from random process theory that
$\frac{h_{rms}^{\prime\prime}}{h_{rms}^{\prime}}\simeq\sqrt{\frac{3}{4}%
\frac{1-H}{2-H}}\left(  \frac{2\pi}{\lambda_{s}}\right)  \simeq\theta\frac
{\pi}{\lambda_{s}}$, where $\lambda_{s}$ is the smallest wavelength of
roughness, $H$ is the Hurst exponent, and $\theta=0.707,1.0$,$1.11$,
respectively for $H=0.3-0.5-0.8$. Therefore, geometry fixes $d_{rep}$
independently on load, and also the gap function is therefore only dependent
on geometry.

If the contact areas were circular, applying Hertz theory would result in a
constant compression $\delta=d_{rep}^{2}/R=4h_{rms}^{\prime2}/h_{rms}%
^{\prime\prime}$, or else%
\begin{equation}
\delta\simeq\frac{4}{\pi\theta}h_{rms}^{\prime}\lambda_{s}.
\end{equation}

In their simplified DMT calculation using first order term for the gap
we\ estimated (Ciavarella, 2016)%
\begin{equation}
\frac{P_{att}}{2\pi wR}=\frac{3^{2/3}}{2}\left(  \frac{\delta}{\Delta
r}\right)  ^{1/3}%
\end{equation}
Finally, as $\Delta r=1.1a_{0}$, for their case of the attractive potential
$l_{a}/a_{0}=0.05\ $\ ($a_{0}$ is atomic distance), and $\frac{\lambda_{s}%
}{a_{0}}=4,8,16,32,64$,
\begin{align*}
\frac{P_{att}}{2\pi wR} &  \simeq\frac{3^{2/3}}{2}\left(  \frac{4}{\pi\theta
}h_{rms}^{\prime}\frac{\lambda_{s}}{a_{0}}\right)  ^{1/3}\\
&  \simeq0.83...2.08\text{ (}h_{rms}^{\prime}=0.1\text{)}\\
&  \simeq1.2...3.02\text{ (}h_{rms}^{\prime}=0.3\text{)}%
\end{align*}
which is clearly up to 2 or 3 times higher than the value expected by the
Maugis limit for $\mu=0$ (Maugis, 2000) -- which in turn means the error is
probably much larger than 2 or 3. \ In other words, this suggests their
contacts \textit{cannot be circular}.

It remains to see therefore an estimate of a line contact geometry. Then, from
our DMT estimate with $\beta=1$ (only first term in the gap function)%
\[
P_{DMT,att}^{\ast}\simeq-\left(  \frac{\mu}{a^{\ast}}\right)  ^{1/3}%
\]

However, the dimensionless contact areas (\ref{adim}) is, using again
(\ref{drep}), Tabor parameter (\ref{Tabor}), and $R=2/h_{rms}^{\prime\prime}$%
\begin{equation}
a^{\ast}=\frac{h_{rms}^{\prime}}{2\pi^{1/3}}\frac{\mu E}{\sigma_{th}}%
\end{equation}
and hence, the PR-DMT 2D model becomes independent on Tabor, but only on slope%
\begin{equation}
P_{PR,att}^{\ast}=-\left(  2\pi^{1/3}\frac{\sigma_{th}}{h_{rms}^{\prime}%
E}\right)  ^{1/3}%
\end{equation}
which clearly becomes arbitrarily large at small slopes.

The real total load curves for PR are therefore (from their potential, we find
$\frac{E}{\sigma_{th}}\simeq\frac{1}{0.07}=\allowbreak14.3$ for$\frac{l_{a}%
}{a_{0}}=0.05$)
\[
P_{PR}^{\ast}=P_{Hertz}^{\ast}-0.59\frac{1}{h_{rms}^{\prime1/3}}\text{ }%
\]
We estimate the Tabor parameter at small scale for the PR data is about
0.4-0.5 for $\lambda_{s}/a_{0}=4$, and about 1 for $\lambda_{s}/a_{0}=64$,
when $h_{rms}^{\prime}=0.1$. For $h_{rms}^{\prime}=0.3$ values are smaller by
a factor close to 2 (0.27 and 0.67). This is for $l_{a}/a_{0}=0.05$ whereas
for $l_{a}/a_{0}=0.005$, they are slightly larger. However, notice that the
DMT-PR estimate does not depend on $\lambda_{s}/a_{0}$ wavelength, whereas the
Maugis solution does, and therefore this trend is not captured with the simple
DMT estimate.

Therefore, for $l_{a}/a_{0}=0.05$ we report in Fig.5a the case $h_{rms}%
^{\prime}=0.1\ $together with Maugis with $\mu=0.5,1$, whereas in Fig.5b,\ the
case $h_{rms}^{\prime}=0.3\ $together with Maugis with $\mu=0.25,0.5.$ It is
clear that the agreement is reasonable and may explain the fit done by PR. In
particular, the lower Tabor parameter seem to be better fitted by the simple
DMT-PR law. In the PR results, these are the low $\lambda_{s}/a_{0}$ which are
the stickiest cases. For the high $\lambda_{s}/a_{0}$, we expect that DMT-PR
underestimates the force of attraction, and therefore surfaces are stickier
than the PR criterion would suggest. This seems in agreement with the PR
findings for pull-off, which do not fit their stickiness criterion (see
Ciavarella, 2016).

However, more investigation is needed to understand within which limit one can
use this approximation for rough contacts. For very small slopes, PR-DMT would
tend to extremely large attractive forces without limit, whereas when the
Tabor parameter increases to values larger than 5, the correct adhesion
solution becomes JKR and doesn't show an unbounded attractive force.

\begin{center}%
\begin{tabular}
[c]{ll}%
{\includegraphics[
height=3.3572in,
width=5.0548in
]%
{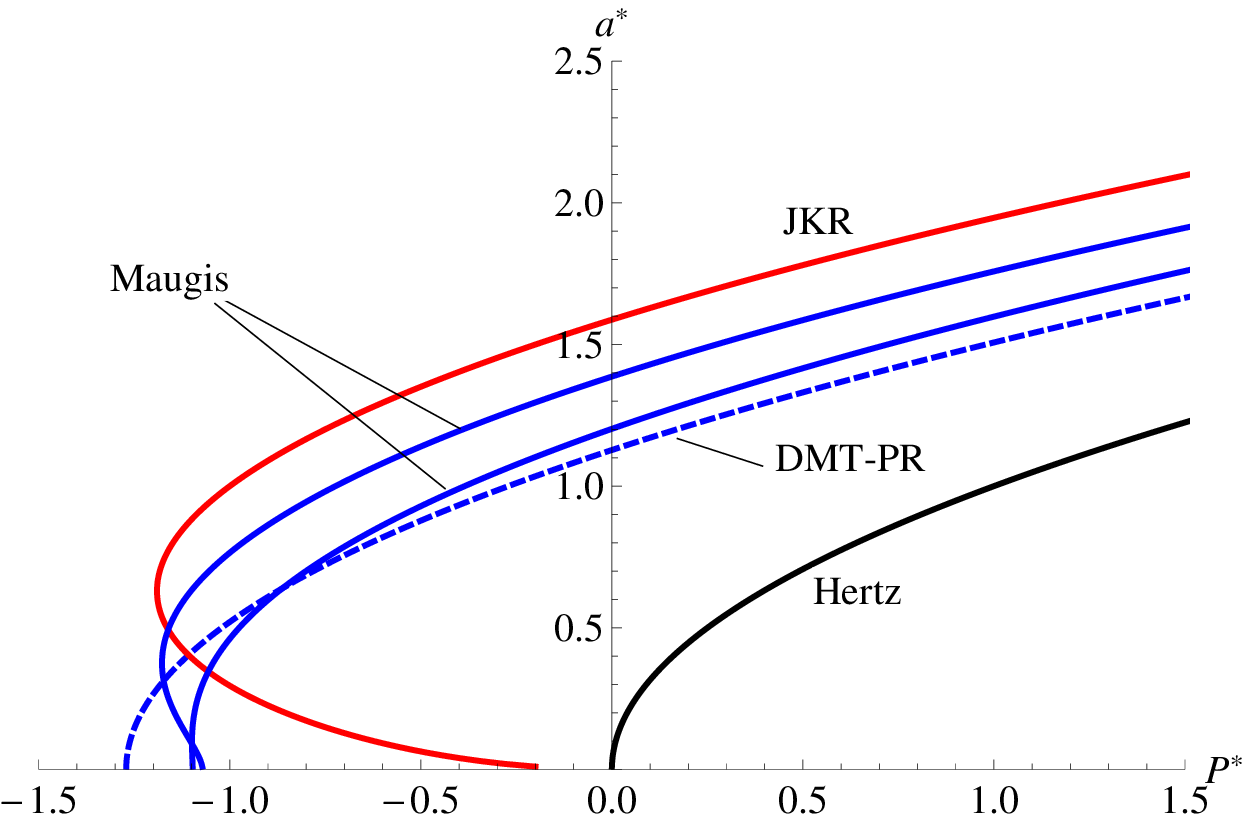}%
}
& (a)\\%
{\includegraphics[
height=3.3572in,
width=5.0548in
]%
{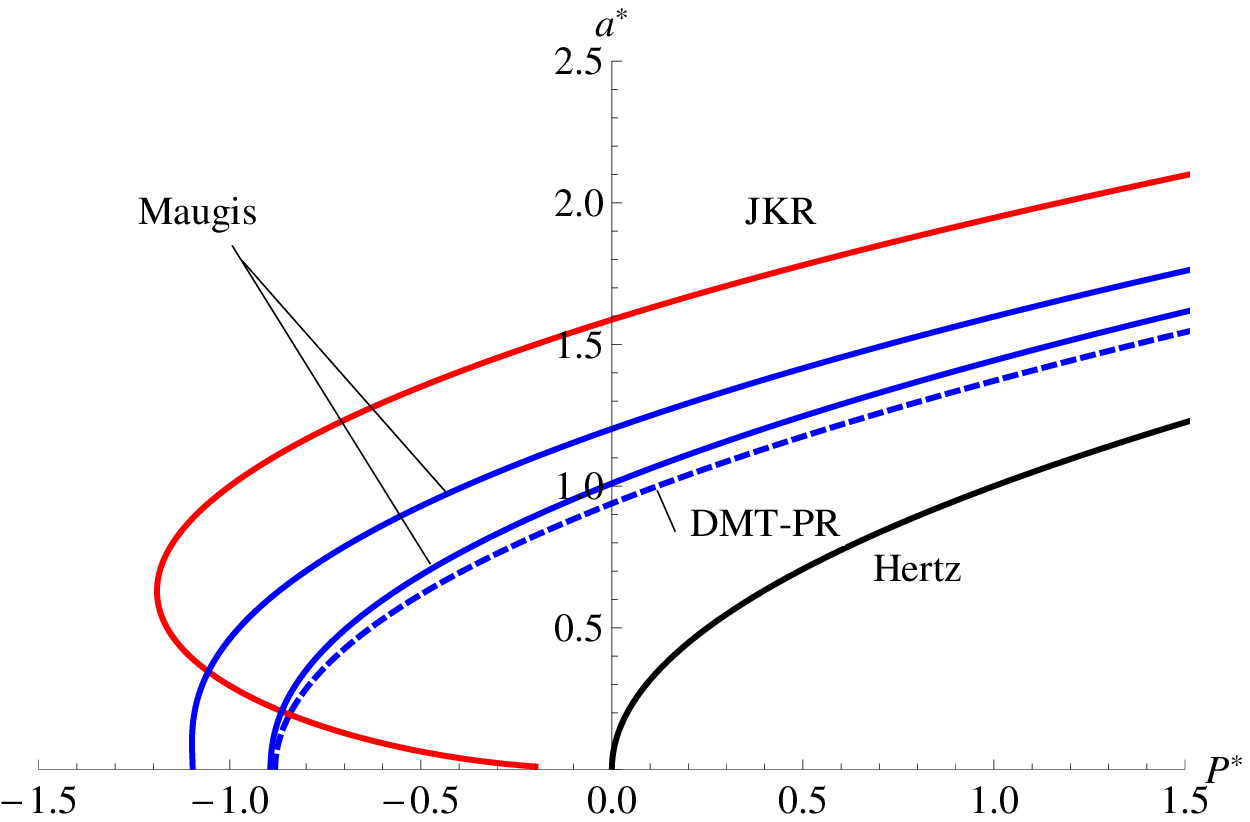}%
}
& (b)
\end{tabular}

Fig.5. Line contact vs. rough contact in PR numerical simulations. Comparison
of Maugis "exact" solution (solid blue line) for estimated range of Tabor
parameter, with the DMT-PR model with first order term of gap (dashed line)
for $h_{rms}^{\prime}=0.1\ (a)$, and $h_{rms}^{\prime}=0.3\ (b).$ In both
figures, $l_{a}/a_{0}=0.05$

\bigskip
\end{center}

\section{Conclusion}

We have examined in details the effect of the DMT approximations ("force
method")\ in the case of a spherical and line contact, and we have shown that:

\begin{itemize}
\item for spherical contact, DMT leads to an error overestimating the
attractive forces, which is very significant already at very low Tabor
parameters, such as $\mu=0.05,$ but becomes extremely large at Tabor of the
order of 1 --- much larger than the error for the pull-off value, which occurs
for a wrong value of separation;

\item for line contact, DMT leads to a different error: the attractive force
tends to be underestimated, except near pull-off, where now the error can be
very significant at $\mu=1$

\item it could be that for elongated contacts, like those found in rough
contact, the error could be intermediate between these two extreme conditions,

\item in general, it is highly unlikely that DMT can be a good approximation
for values of Tabor parameter greater than, say, 0.25.

\item in rough contacts, due to special features of the contact area (far from
set of spherical asperities like in the classical Fuller-Tabor (1975) model
based on Nayak (1971) geometries), the contact appears a set of elongated
contacts, whose representative diameter is relatively independent on load ---
but this is probably only true during loading.

\item the use of only asymptotic term of the gap function leads to serious
errors, however when adding the further crude approximation made by Pastewka
\&\ Robbins that the slope at the contact edge was given by geometrical
considerations only, it leads to some more reasonable behaviour. However,
quantitatively the fit they may have obtained is only a pure coincidence

\item the scaling equations obtained from DMT type of model by Pastewka and
Robbins should not be used outside their range of parameters, as extrapolation
is not likely to work. In particular, for very small rms slopes of surface,
the limit results do not seem realistic. Also, the dependence on Tabor
parameter is ignored for the data with different short wavelengths than in the
PR surfaces.
\end{itemize}

\section{References}

\bigskip R.S. Bradley, (1932) The cohesive force between solid surfaces and
the surface energy of solids. Phil. Mag. 13, 853.

M. Ciavarella, (2015). Adhesive rough contacts near complete contact.
International Journal of Mechanical Sciences, 104, 104-111.

M. Ciavarella, (2016). On a recent stickiness criterion using a very simple
generalization of DMT theory of adhesion. Journal of Adhesion Science and
Technology, 30(24), 2725-2735.

B.V. Derjaguin, (1934) Theorie des Anhaftens kleiner Teilchen. Kolloid
Zeitschrift 69, 155.

\bigskip\bigskip B.V. Derjaguin, V.M. Muller and Yu.P. Toporov, (1975) Effect
of contact deformations on the adhesion of particles. J. Colloid Interface
Sci. 53, 314.

K. N. G. Fuller, \& D.Tabor, (1975). The effect of surface roughness on the
adhesion of elastic solids. In Proceedings of the Royal Society of London A:
Mathematical, Physical and Engineering Sciences (Vol. 345, No. 1642, pp. 327-342).

JA. Greenwood, (2007). On the DMT theory. Tribology Letters, 26(3), 203-211.

F. Jin, W. Zhang, S. Zhang, X. Guo, (2014). Adhesion between elastic cylinders
based on the double-Hertz model. International Journal of Solids and
Structures, 51(14), 2706-2712.

K.L. Johnson, J.A. Greenwood, (2008). A Maugis analysis of adhesive line
contact. J. Phys. D: Appl. Phys. 41, 155315-1--155315-6.

V.M. Muller, V.S. Yuschenko and B.V. Derjaguin, (1980). On the influence of
molecular forces on the deformation of an elastic sphere and its sticking to a
rigid plane. J. Colloid Interface Sci. 77, 91.

V.M. Muller, B.V. Derjaguin and Yu.P. Toporov, (1983) On two methods of
calculation of the force of sticking of an elastic sphere to a rigid plane.
Colloids Surf 7. 251.

M.D. Pashley, (1984). Further consideration of the DMT model for elastic
contact. Colloids Surf 12, 69.

KL Johnson, K. Kendall, and A. D. Roberts. (1971). Surface energy and the
contact of elastic solids. Proc Royal Soc London A: 324. 1558.

D Maugis, (2000). Contact, adhesion and rupture of elastic solids (Vol. 130).
Springer, New York.

C.A. Morrow, M.R. Lovell, (2005). An extension to a cohesive zone solution for
adhesive cylinders. J. Tribol. 127, 447--450.

V.M. Muller, V.S. Yuschenko and B.V. Derjaguin, (1980). On the influence of
molecular forces on the deformation of an elastic sphere and its sticking to a
rigid plane. J. Colloid Interface Sci. 77, 91.

V.M. Muller, B.V. Derjaguin and Yu.P. Toporov, (1983) On two methods of
calculation of the force of sticking of an elastic sphere to a rigid plane.
Colloids Surf 7. 251.

P.R. Nayak, Random process model of rough surfaces in plastic contact, Wear 26
(1973) 305--333.

M. D.Pashley, (1984). Further consideration of the DMT model for elastic
contact. Colloids and surfaces, 12, 69-77.

L. Pastewka, \& M.O.\ Robbins, (2014). Contact between rough surfaces and a
criterion for macroscopic adhesion. Proceedings of the National Academy of
Sciences, 111(9), 3298-3303.

B.N.J. Persson, (2002). Adhesion between an elastic body and a randomly rough
hard surface, Eur. Phys. J. E 8, 385--401

B. N. Persson, \& M. Scaraggi, (2014). Theory of adhesion: Role of surface
roughness. The Journal of chemical physics, 141(12), 124701.

D. Tabor, (1977) Surface forces and surface interactions. J. Colloid Interface
Sci. 58, 2.

\bigskip

\end{document}